\documentclass[nofootinbib,twocolumn,preprintnumbers]{revtex4-1}
\usepackage[dvips]{graphicx}
\usepackage{amsmath,amsthm,amssymb}
\begin{document}

\topmargin -1.5 cm

\def\lsim{\mathrel{\rlap{\lower4pt\hbox{\hskip1pt$\sim$}}
    \raise1pt\hbox{$<$}}}
\def\gsim{\mathrel{\rlap{\lower4pt\hbox{\hskip1pt$\sim$}}
    \raise1pt\hbox{$>$}}}
\newcommand{\vev}[1]{ \left\langle {#1} \right\rangle }
\newcommand{\bra}[1]{ \langle {#1} | }
\newcommand{\ket}[1]{ | {#1} \rangle }
\newcommand{\ev}{ {\rm eV} }
\newcommand{\kev}{{\rm keV}}
\newcommand{\mev}{{\rm MeV}}
\newcommand{\gev}{{\rm GeV}}
\newcommand{\tev}{{\rm TeV}}
\newcommand{\mpl}{$M_{Pl}$}
\newcommand{\mw}{$M_{W}$}
\newcommand{\Ft}{F_{T}}
\newcommand{\Zparity}{\mathbb{Z}_2}
\newcommand{\BLambda}{\boldsymbol{\lambda}}
\newcommand{\be}{\begin{eqnarray}}
\newcommand{\ee}{\end{eqnarray}}
\newcommand{\Feyn}[1]{#1\kern-0.60em/}

\title{Associated Production of Non-Standard Higgs Bosons at the LHC}
\author{David E. Kaplan, Matthew McEvoy}
\affiliation{Department of Physics and Astronomy,
		Johns Hopkins University,
		Baltimore, MD  21218}
\date{\today}
\begin{abstract}
We discuss the feasibility of seeing a Higgs boson which decays to four partons through a pair of (pseudo-)scalars at the LHC.  We restrict our search to Higgs bosons produced in association with a $W/Z$ boson at high transverse momentum.  We argue that subjet analysis techniques are a good discriminant between such events and $W/Z$ plus jets and $t\bar{t}$ production. For light scalar masses (below $30$ GeV), we find evidence that a flavor-independent search for such a non-standard Higgs boson is plausible with $100$ fb$^{-1}$ of data, while a Higgs decaying to heavier scalars is only likely to be visible in models where scalar decays to $b$ quarks dominate.   
\end{abstract}

\maketitle

\section{Introduction}

Significant effort is being made to discover the Higgs boson at the Large Hadron Collider.  If the Standard Model (SM) Higgs boson is ruled out in all of its allowed mass range, the next step will be to find out if a Higgs (a remnant of weakly coupled electroweak symmetry breaking) exists at all.  In doing so, one must consider all possible reasonable decay modes due to new physics.  

The phenomenology of a SM Higgs boson with mass $m_h \lsim 135$ GeV is dominated by the relatively weak Yukawa couplings characterized by the ratio of light fermion masses and the vacuum expectation value (vev) of the Higgs. Because the Higgs vev ($\sim 246$ GeV) is so much larger than the mass of any of the fermions to which the Higgs boson can decay, given restriction on $m_h$ from precision electroweak tests \cite{ewwg}, these couplings are quite small, and therefore the width of the light Higgs boson into SM particles is quite low ($\sim 3$ MeV for $m_h = 120$ GeV). With such a small decay width, new light particles, even with a relatively small couplings to the Higgs, could easily be its dominate decay product. Under any such model, SM Higgs searches could be rendered useless. 

The simplest possible extension to the Higgs sector involves the addition of a gauge singlet scalar, $a$. Various new physics models contain such a field, such as non-minimal supersymmetry and composite (and Little) Higgs models (for a review, see \cite{Chang:2008cw}). Given a coupling between the Higgs field and the $a$ field $\lambda_a H^2 a^2$ electroweak symmetry breaking allows the decay $h \rightarrow aa$. Without any additional terms in the Lagrangian, the $a$ would be stable and experimentalists would need to search for an invisible Higgs -- such a phenomenological possibility has been studied in, e.g. \cite{LHCInvH}.  Other reasonable possibilities include a model in which the $a$ and the Higgs boson mix, in which case the $a$ picks up SM Higgs branching ratios to SM particles. For most of the reasonable parameter space, this would lead to the dominant decay chain $h \rightarrow 2a \rightarrow 4b$.  Searches for this final state have been performed with LEP data and a lower bound of 110 GeV has been set \cite{Schael:2006cr}.  Other simple and natural models, where $a$ is a pseudo-Goldstone boson \cite{Lisanti:2009vy}, can easily allow $a\rightarrow gg$ (gluons) to dominate (see for example \cite{Chang:2005ht}) or lighter quarks (see \cite{Bellazzini:2009kw}).  

Previous work has been done on the phenomenology of $h\rightarrow 4b$ at the LHC \cite{Ellwanger:2003jt, Carena:2007jk}.  The backgrounds for such signals are large, as are those for the SM $h\rightarrow b\bar{b}$.  The SM case was approached more recently by looking at the phase space where the Higgs is produced in association with a vector boson, and the two particles are produced with large $p_T$ \cite{Butterworth:2008iy}.  Not only are leptonic decays of the gauge boson a nice discriminant from QCD production, the SM background ($W$ and $Z$ plus jets) drops much faster than the signal at large $p_T$ due to the fact that the background is dominated by t-channel diagrams while the signal is dominated by the s-channel.  Their analysis used jet deconstruction and reconstruction tools partly developed in early work \cite{Butterworth:2002tt}.  The analysis of the boosted Higgs \cite{Butterworth:2008iy} showed that a once hopeless signal became hopeful.  

In this paper, we discuss the application of boosted Higgses analyses to non-standard Higgs decays can have a dramatic impact on the reach of the LHC, not only in the $h\rightarrow 4b$ channel, but in the notorious $h\rightarrow 4g$ channel as well.  Vector boson plus jets and 
$t\bar{t}$ production will provide significant backgrounds to our signal, but we find that discovery significance can be achieved in both channels for significant ranges of $a$ masses with more than 100 fb$^{-1}$ of data.  Work along these lines with much smaller ranges of parameters studied \cite{Chen:2010wk, Falkowski:2010hi}, or with significant model dependence \cite{Bellazzini:2010uk}, have appeared previously.  This work had previously appeared in a PhD. thesis \cite{mcevoy-thesis}.  



\section{Analysis and Results}

We generate both signals and backgrounds using the MadGraph/MadEvent v4.3 package \cite{MadGraph} \cite{MadEvent}. These events are showered through Pythia \cite{Sjostrand:2006za} and jet clustering is performed using the routines in the FastJet \cite{FastJet} libraries.  

Vector bosons which decay hadronically are difficult to distinguish from hadronic activity arising through QCD, so we proceed based on an analysis of leptonic decays of vector bosons. These come in three types for our signal: $Zh \rightarrow \ell^{+}\ell^{-}h$, $Zh \rightarrow \Feyn{E}_T h$, and $W^{\pm}h \rightarrow \ell^{\pm}\Feyn{E}_T h$, where $\ell^{\pm}$ represents a charged light lepton (muon or electron).  In order to simulate lepton isolation cuts applied experimentally, we neglect from our analysis leptons which have $p_T \leq 10$ GeV, or for which there is significant hadronic activity in an isolation cone of size $\Delta R = 0.4$ around the lepton (significant meaning a scalar sum of $p_T$ more than $20\%$ of the $p_T$ of the lepton itself). Leptons vetoed in this way are included in the hadronic activity of the event.  

Events are then classified into one of three channels; the first requires the presence of two opposite-charge, same flavor leptons which reconstruct to an invariant mass of $91\pm 10$ GeV with total $p_T \geq 200$ GeV.  Failing this, events with a lepton and missing $E_T$ which reconstruct to an invariant transverse mass of at most $90$ GeV ($10$ GeV greater than the $W$ mass) with a total $p_T \geq 200$ GeV are accepted into the second channel.  Finally, events which simply have $\Feyn{E}_T \geq 200$ GeV are accepted via the third channel.  

The energy resolution on electrons and muons at ATLAS and CMS is expected to be of the order of a few percent, while the resolution on missing $E_T$ is expected to be $\lsim 10$ GeV for events in which the scalar sum of $p_T$ is $\lsim 500$ GeV. We neglect the energy and angular resolution on isolated light leptons. The production cross-sections $pp \rightarrow Vh$, $pp \rightarrow V+jets$, and $pp \rightarrow t\bar{t} \rightarrow W^{+}bW^{-}\bar{b}$ are relatively flat as a function of the $p_T$ of the outgoing vector boson ($\sim10\%$ difference from $190-200$ GeV, and from $200-210$ GeV).  The effect of smearing is to change the number of events that pass the cuts by at most a few percent, and thus we ignore the missing $E_T$ smearing.  

As can be seen from Table \ref{tab:lepcuts}, the efficiency with which signal events (in which the boson decays leptonically or invisibly and has sufficient $p_T$) pass these cuts is $\sim 75\%$, and most of the passed events are characterized properly (e.g. events with a $W$ boson pass the lepton + missing $E_T$ cut). However, a significant minority of $t\bar{t}$ pass the pure $\Feyn{E}_T$ cut instead of $\ell+\Feyn{E}_T$, due to the charged lepton being non-central or too close to a hadronic jet.  

\begin{table}
\caption{Cross-sections in picobarns after successive cuts on leptons and $\Feyn{E}_T$ for associated Higgs production and backgrounds. The ``Inclusive'' column gives the production cross-section with no cuts. The ``$V\: p_T$'' column gives the production cross-section for events in which there is a vector boson ($W$ or $Z$) with $p_T \geq 200$ GeV (including all decay modes for this boson). These cross-sections are presented to leading order.}
\begin{center}
\begin{tabular}{|c|c|c|c|c|c|}
\hline
Process&Inclusive&$V\: p_T$&$\ell^+ \ell^-$& $\ell +\Feyn{E}_T$&$\Feyn{E}_T$\\
\hline
$Wh$&$0.87$&$0.047$&$<0.001$&$0.011$&$<0.001$\\
\hline
$Zh$&$0.74$&$0.038$&$0.0018$&$<0.001$&$0.0063$\\
\hline
$W+jets$&$28600$&$180$&$<0.1$&$30.8$&$1.8$\\
\hline
$Z+jets$&$9300$&$80.6$&$3.0$&$0.5$&$15.5$\\
\hline
$t\bar{t}$&$610$&$54.3$&$<0.1$&$8.4$&$2.2$\\
\hline
\end{tabular}
\end{center}
\label{tab:lepcuts}
\end{table}  

Once the cuts on reconstructed bosons have been applied, we proceed to an analysis of the remaining outgoing energy (we will refer to this as the ``hadronic'' side of the event, though it includes outgoing particles which will be captured by the EM and hadronic calorimeters as well as the muon chambers). The calorimeters at the LHC have finite angular granularity on the scale of $\Delta \phi \sim \Delta \eta < 0.1$ in most of the acceptance region \cite{AngularResolution}. As we will be doing an analysis of the substructure of jets on a scale not too much larger than this, we simulate it by binning hadronic energy into $63$ equally-sized bins in azimuthal angle $\phi$ and into $100$ equally-sized bins in pseudorapidity (for $-5\leq\eta\leq 5$. We do not simulate the finite energy resolution of the EM and Hadron calorimeters themselves, but we expect this effect to be subdominant to the effects of the jet algorithms and the finite granularity. The energy resolution is expected to be of the order of $10\%$ for the relevant energy scales \cite{CMScal} while the cuts we will perform assume resolutions somewhat worse than this.

Once this binning has been accomplished the resulting ``particles'' (representing energy deposits in the calorimeter) are clustered according the Cambridge-Aachen jet algorithm with a merging cutoff parameter of $R_{max} = 1.2$ \cite{Dokshitzer:1997in}. This value of $R_{max}$ was optimized empirically for a light Higgs mass $m_h = 120$ GeV in order to capture the decay products of the Higgs boson in a single jet. 
Jets are only kept if the clustered jet has $p_T \geq 30$ GeV.

The Cambridge-Aachen jet algorithm has a useful property; as it is a monotonic iterative procedure (at every step objects can only be merged, never unmerged) it is possible to work backwards from a single final jet to study its component subjets. The method by which we accomplish this is similar to the that used in the study of boosted top quarks in \cite{BoostedTops}. The candidate Higgs jet (taken to be the highest $p_T$ jet in each event) is iteratively decomposed. At every iteration every (sub)jet is examined to determine if it can be broken down further according to two criteria: that the two components are separated by $\Delta R \geq R_{min}$ and each has $p_T \geq \zeta p_{T}^h$ where $p_{T}^h$ is the transverse momentum of the original Higgs candidate jet and $R_{min}$ and $\zeta$ are two input parameters, set to $0.31$ and $0.2$ respectively in this analysis ($R_{min}$ is chosen so that it is not an exact multiple of cell spacing). The results we obtain are not particularly sensitive to these choices for a range of parameters $0.2 \leq R_{min} \leq 0.5$ and $0.15 \leq \zeta \leq 0.3$.  

\begin{figure}
  \resizebox{\hsize}{!}{\rotatebox{-90}{\includegraphics{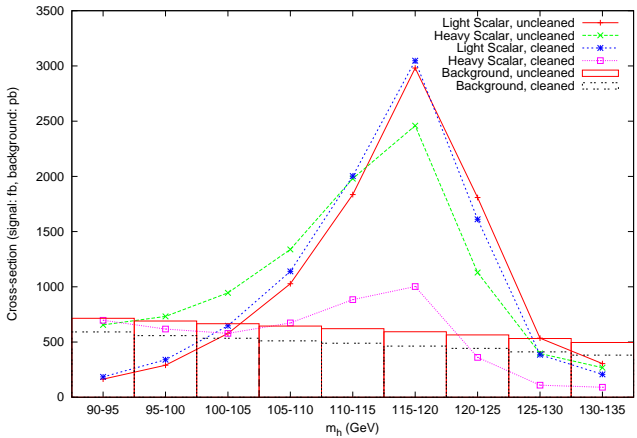}}}
  \caption{Cross-sections (per $5$ GeV bin) versus the reconstructed Higgs mass. Cleaned masses have been broken into subjets and reconstructed from these. In the plot, ``light scalars'' have a mass of $20$ GeV and ``heavy scalars'', $50$ GeV. The Higgs mass used to produce the signal was $120$ GeV. Signals and backgrounds represent the sum of the three reconstructed vector boson channels. The data presented are for the $h \rightarrow a a \rightarrow b\bar{b}b\bar{b}$ model. } 
\label{fig:cleanvunclean}
\end{figure}

The examination of every (sub)jet is itself iterative; if the two immediate parents of a (sub)jet do not pass the above criteria then the leading parent (in $p_T$) is itself examined to determine if it can be broken in two, and so forth, while the softer parent is removed from consideration. If this procedure reaches the level of primitive objects (where the objects have no parents) then the original (sub)jet being examined is accepted as being indivisible. If, on the other hand, the procedure eventually finds two acceptable subjets then the original (sub)jet is broken into these two objects. This can mean that some of the original (sub)jet's energy being lost as rejected parents are removed from the final objects -- we describe this removal as ``cleaning'' the jet, (essentially identical to ``pruning'', introduced in \cite{Ellis:2009su}).  Cleaning affects relatively soft objects well-separated (in $\Delta R$) from most of the energy in the (sub)jet. In Figure \ref{fig:cleanvunclean} we present the difference in reconstructed Higgs mass between the uncleaned mass reconstructed before breaking the Higgs jet into subjets and the cleaned mass reconstructed from subjets.  If the candidate Higgs does not split into two, three or four subjets via the procedure outlined above, then the event is rejected, as our goal is to use jet substructure as a discriminant.  

The phenomenology of $h \rightarrow 2a \rightarrow 4j$ can be thought of as splitting into two disparate regimes: one where the intermediate scalar is relatively light ($m_a \lsim 30$ GeV) and one where it's relatively heavy. In the former case, the intermediate scalars can both be reconstructed fairly easily with high efficiency, while in the latter such a full reconstruction of the two stage decay isn't possible. We therefore proceed with two parallel analyses.  
\\
\subsection{Light Scalars}

Figure \ref{fig:cleanvunclean} demonstrates that the mass resolution on the reconstructed Higgs is relatively unaffected, for signal events with a light intermediate scalar, by the choice of cleaned versus uncleaned mass. On the other hand, the background is reduced by $\sim 20\%$ when using the cleaned mass. Since the analysis detailed in this subsection won't provide a statistically-significant signal at the LHC given a heavy intermediate scalar, we optimize for light scalars by using the cleaned Higgs mass as our first discriminant.  

We require the search to be performed with a specific Higgs mass in mind, and then scan over possible Higgs masses.  An appropriate search would be for Higgs masses between $100$ and $150$ GeV -- the bottom of the range is beyond a reasonable guess of LEP sensitivity to Higgs decays to four gluons \cite{:2001yb}, while the top is set by the mass where Higgs to gauge bosons will dominate over a weakly coupled scalar.  Fine-tuning considerations for the light scalar brings the upper limit of this range to $130$ GeV.  A scan over Higgs masses can be relatively course (every $5$ GeV) considering the energy resolutions allowed by detectors and the jet finding algorithm.  As an example, we choose a Higgs mass of $120$ GeV (both for the Monte Carlo sample and for the simulated search), for which we accept events with a reconstructed mass between $100$ and $125$ GeV.  

If the candidate Higgs jet splits into two subjets then each of these is assumed to arise from a single intermediate scalar. The invariant mass of each of the two jets are compared, and the lighter jet is required to be at least $75\%$ the mass of the heavier jet.  If the candidate Higgs splits into three subjets then the four-momenta of the two lightest subjets are combined and the invariant mass of this object is compared to the mass of the heaviest jet.  Again, the lighter candidate scalar is required to have $75\%$ of the mass of the heavier candidate. Finally, if the candidate Higgs splits into four subjets, then each of the three possible pairings of jets into candidate scalars is attempted and the combination with the closest match in mass is assumed to be correct. The same $75\%$ criterion is applied in this case.  

In Table \ref{tab:hcuts}, we present the effects of the cuts.  First we present the affect on the effective cross section after the combination of the Higgs invariant mass cut and the $\geq 2$ subjets cut is applied.  Second, we present the result of the subsequent cuts on the invariant masses of the intermediate scalars.  The results are shown for each of the three samples -- dileptons, single lepton with missing transverse energy, and only missing transverse energy.  

\begin{table}
\caption{Cross-sections (in femtobarns) after various cuts for associated Higgs production and backgrounds.  Signal events with the decays $h \rightarrow 4b$ and $h \rightarrow 4g$ are presented with a range of $a$ masses (in GeV).  The $m_h$ columns are after a cut requiring that the candidate Higgs jet splits into at least two subjets and has cleaned mass between $100$ and $125$ GeV. The $\Delta m_a$ columns are after a cut requiring that the reconstructed $a$ masses are within $25\%$ of each other. The three pairs of columns are for, respectively, $i)$ the dilepton sample, $ii)$ the $\ell+\Feyn{E}_T$ sample, and $iii)$ the pure $\Feyn{E}_T$ sample. }  
\begin{center}
\begin{tabular}{|c||c|c||c|c||c|c|}
\hline
process&$m_h $&$\Delta m_a $&$m_h $&$\Delta m_a $&$m_h $&$\Delta m_a $\\
\hline
$V+j$   &$98.2$&$24.3$&$990.8$&$244.7$&$511.1$   &$126.3$\\
\hline
$t\bar{t}$   &$0.2$&$<0.1$&$685.9$&$164.7$&$154.3$   &$37.2$\\
\hline
$4b (15)$   &$0.73$&$0.49$&$4.59$&$3.08$&$2.74$   &$1.85$\\
\hline
$4b (20)$   &$0.77$&$0.54$&$4.83$&$3.33$&$2.83$    &$1.97$\\
\hline
$4b (30)$   &$0.68$&$0.36$&$4.28$&$2.21$&$2.52$   &$1.31$\\
\hline
$4b (40)$   &$0.38$&$0.06$&$2.48$&$0.37$&$1.47$   &$0.20$\\
\hline
$4g (15)$   &$0.76$&$0.56$&$4.50$&$3.32$&$2.79$   &$1.99$\\
\hline
$4g (20)$   &$0.80$&$0.61$&$4.78$&$3.65$&$2.85$   &$2.17$\\
\hline
$4g (30)$   &$0.81$&$0.46$&$4.80$&$2.65$&$2.82$   &$1.56$\\
\hline
$4g (40)$   &$0.52$&$0.12$&$3.13$&$0.60$&$1.86$   &$0.35$\\
\hline
\end{tabular}
\end{center}
\label{tab:hcuts}
\end{table}

\begin{figure}
  \resizebox{\hsize}{!}{\rotatebox{-90}{\includegraphics{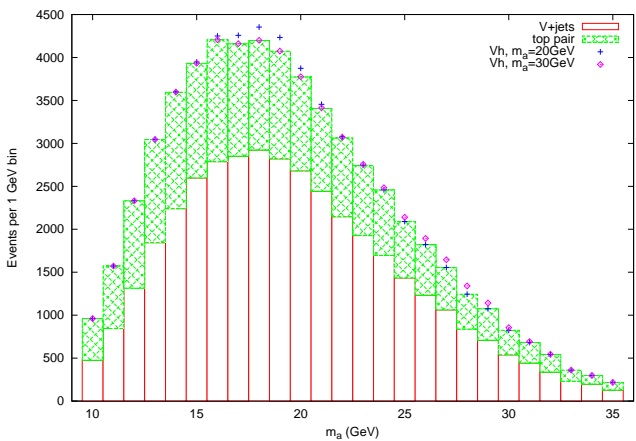}}}
  \caption{Expected events (per $1$ GeV bin) versus the reconstructed intermediate scalar mass. Two different signal cases are presented, with $m_a = 20$ GeV and $m_a = 30$ GeV. Both sets of points represent a model where $a \rightarrow gg$ is the only decay channel. A model in which $a \rightarrow b\bar{b}$ is dominant would give $\sim 10\%$ less signal for the $m_a = 20$ GeV case and $\sim 25\%$ less signal for the $m_a = 30$ GeV case. }  
\label{fig:amass}
\end{figure}

The mean of the masses of the two candidate scalar objects is then used as our final discriminant. As can be seen from Figure \ref{fig:amass} a mass window of $\sim 20 - 30\%$ of the central value is reasonably optimal to maximize the expected significance of the signal.  In Table \ref{tab:lightresults}, we present the expected number of events for various signal cases as well as from the major backgrounds falling in the peak mass window for each signal case.  We also show the expected statistical significance calculated simply as ${\rm signal}/\sqrt{\rm background}$. The results are presented as the sum across all three channels, $(i)$, $(ii)$, and $(iii)$. All of our cuts in this analysis are identical across the three channels, and none of the channels appears to be significantly favored relative to any other channel by the cuts applied on the candidate Higgs jet. We obtain $\sim 3.5 \sigma$ for $m_a \leq 30$ GeV. Above this threshold it becomes more and more unlikely to properly reconstruct the intermediate scalars through the technique above, and the drop-off in expected significance is severe. We present results for a model with a branching ratio of $1$ for $a \rightarrow b\bar{b}$ as well as a model with a branching ratio of $1$ for $a \rightarrow gg$. The results are broadly similar, with $a \rightarrow gg$ giving somewhat better reconstruction efficiencies. This suggests that the preceding type of analysis provides significant discovery potential at the LHC for a large range of intermediate scalar masses and any model in which the intermediate scalar preferentially decays into pairs of colored objects (even if the decay width of the intermediate scalar is shared between several channels). Such model independence is a powerful reason to prefer this analysis over, for example, an analysis which depends on the flavor-tagging of decay products.  

The data presented in Figure \ref{fig:amass} represent only part of the information which would be available to an experimentalist pursuing a similar analysis. In order to understand experimental backgrounds more fully than is possible merely through Monte Carlo simulation, there are a number of physical handles which can provide information about the individual processes which make up the background. For example, the $l^{+}l^{-}$ channel $(i)$ has almost no $t\bar{t}$ component, while the $t\bar{t}$ background sample can be made more pure by making an extra jet requirement. A further possibility is to look at $\gamma + jets$ events. This is a purely SM background sample; the topology of the diagrams which contribute to such production is the same as for those which contribute to the $V + jets$ background. Only the mass of the vector boson and a single coupling constant differ. Analysis of the hadronic components of $\gamma + jets$ (or Drell-Yan plus jet outside of the $Z$-mass window) events could yield substantial understanding of the hadronic components of $V + jets$.  

\begin{table}
\caption{Expected number of events (with 100 fb$^{-1}$ of data) lying under the peak signal reconstructed mass window (in GeV) for various intermediate scalar masses (both for $h \rightarrow 4b$ and $h \rightarrow 4g$), along with statistical significance $\sigma = s/\sqrt{b}$. Note the precipitous drop-off in signal above $m_a = 30$ GeV. } 
\begin{center}
\begin{tabular}{|c|c|c|c|c|c|c|}
\hline
Process&Window&Signal&$V+j$&$t\bar{t}$&$s/b$&$\sigma$\\
\hline
$4b(15)$&$12-17$&$479$&$10780$&$6320$&$2.8\%$&$3.7$\\
\hline
$4b(20)$&$16-22$&$536$&$16500$&$7310$&$2.2\%$&$3.5$\\
\hline
$4b(30)$&$25-31$&$317$&$5810$&$2800$&$3.7\%$&$3.4$\\
\hline
$4b(40)$&$32-40$&$34$&$1130$&$670$&$1.9\%$&$0.8$\\
\hline
$4g(15)$&$12-17$&$523$&$10780$&$6320$&$3.1\%$&$4.0$\\
\hline
$4g(20)$&$16-22$&$608$&$16500$&$7310$&$2.5\%$&$3.9$\\
\hline
$4g(30)$&$25-31$&$420$&$5810$&$2800$&$4.9\%$&$4.5$\\
\hline
$4g(40)$&$32-40$&$65$&$1130$&$670$&$3.6\%$&$1.5$\\
\hline
\end{tabular}
\end{center}
\label{tab:lightresults}
\end{table} %

\begin{figure}
  \resizebox{\hsize}{!}{\rotatebox{-90}{\includegraphics{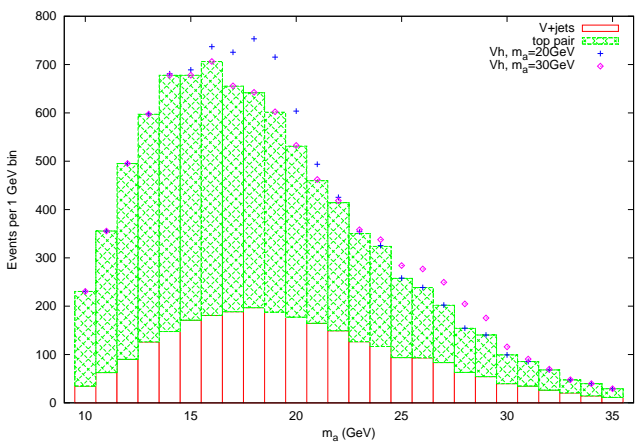}}}
  \caption{Expected events (per $1$ GeV bin) versus the reconstructed intermediate scalar mass. Two different signal cases are presented, with $m_a = 20$ GeV and $m_a = 30$ GeV. Both sets of points represent a model where $a \rightarrow b\bar{b}$ is the only decay channel. We require that at least one of the candidate Higgs subjets receives a $b$ tag. }  
\label{fig:amassbtags}
\end{figure}

While we have presented a search strategy for hadronically decaying $a$ scalars, the dominant $a$ decay in some models is to $b\bar{b}$. For this case, we can ask how the prospects for the search are affected by requiring one or two $b$ tags on the Higgs component subjets. The $b$-tagging probabilities we use are estimated using the method described in the following subsection. Table \ref{tab:lightbresults} shows the effect of requiring $b$ tags on the number of expected signal and background events in the appropriate peak signal region given in Table \ref{tab:lightresults}. In Figure \ref{fig:amassbtags} we reproduce the equivalent of Figure \ref{fig:amass} for the $a \rightarrow b\bar{b}$ model when requiring that at least one subjet receives a $b$ tag.  

\begin{table}
\caption{Expected number of events (with 100 fb$^{-1}$ of data) requiring $1$ and $2$ $b$ tags (first and set set of columns respectfully) for various values of $m_a$ in GeV. We assume a branching ratio of $1$ for $h\rightarrow 4b$. Signals and backgrounds for each $m_a$ are those lying under the mass window given in Table \ref{tab:lightresults}. Background represents the sum of the $V+jets$ background and $t\bar{t}$ background. Also provided is $\sigma = \frac{s}{\sqrt{b}}$.  } 
\begin{center}
\begin{tabular}{|c||c|c|c||c|c|c|}
\hline
$m_a$   & signal &background& $\sigma$ &signal&background&$\sigma$ \\
\hline
$15$   &$414$&$3150$&$7.4$&$191$   &$122$   &$17.3$\\
\hline
$20$   &$433$&$3600$&$7.2$&$167$   &$130$   &$14.7$\\
\hline
$30$   &$215$&$1090$&$6.5$&$64$   &$39$   &$10.3$\\
\hline
$40$   &$21$&$230$&$1.4$&$7$   &$11$   &$2.2$\\
\hline
\end{tabular}
\end{center}
\label{tab:lightbresults}
\end{table}  

\subsection{Heavy Scalars}

As seen above, we cannot rely on reconstructing the intermediate scalars properly when $m_a$ is much above $30$ GeV.  In this section we will discuss the discovery potential for the specific Higgs cascade decay $h \rightarrow aa \rightarrow b\bar{b}b\bar{b}$. The additional discriminant provided by flavor-tagging is necessary to provide statistical significance of the signal with a heavier intermediate scalar.  

As noted previously, for heavier scalars the uncleaned reconstructed Higgs mass is much better resolved than the cleaned Higgs mass, so we make our first cut on the uncleaned mass, passing events in the window $100-125$ GeV. We will be applying a $b$-tagging requirement, so the $\ell + \Feyn{E}_T$ channel (labeled $(ii)$ earlier) will be dominated by $t\bar{t}$ (since virtually every $t$ decay includes a $b$ while only a minority of QCD jets in $V + jets$ are $b$s). The $t\bar{t}$ events have significantly more hadronic activity than do signal or $V + jets$ events. We therefore apply a jet veto -- all events with at least one additional $p_T>30$ GeV jet are rejected -- to reduce the dominant $t\bar{t}$ background.  Finally, the fact that this analysis is aimed at heavier intermediate scalars allows us to require the candidate Higgs jet to split into $3+$ subjets.  The effect of these two cuts is summarized in Table \ref{tab:vetocuts}.  

\begin{table}
\caption{Cross-sections (in fb) after a cut on the uncleaned reconstructed Higgs mass and a jet veto for associated Higgs production and backgrounds. The three sets of columns are for $(i)$ the dilepton channel, $(ii)$ the $\ell+\Feyn{E}_T$ channel and $(iii)$ the pure $\Feyn{E}_T$ channel.  Note, a jet veto is not necessary for channel $(i)$.  The final column is the sum over all channels after requiring at least 3 subjets from the candidate Higgs. }  
\begin{center}
\begin{tabular}{|c||c||c|c||c|c||c|}
\hline
process&$m_h $&$m_h $&Veto&$m_h $&Veto&subjet\\
\hline
$V+j$       &$127.2$&$1282$&$725.3$&$658.5$     &$365.2$&$138.6$\\
\hline
$t\bar{t}$  &$0.3$&$833.3$&$92.6$&$188.3$      &$11.4$&$7.86$\\
\hline
$Vh (15)$  &$0.72$&$4.47$&$3.21$&$2.68$     &$1.92$&$0.05$\\
\hline
$Vh (20)$  &$0.76$&$4.70$&$3.40$&$2.77$     &$1.97$&$0.09$\\
\hline
$Vh (30)$  &$0.71$&$4.39$&$3.13$&$2.59$     &$1.82$&$1.14$\\
\hline
$Vh (40)$  &$0.68$&$4.35$&$3.15$&$2.54$     &$1.80$&$1.81$\\
\hline
$Vh (50)$  &$0.70$&$4.50$&$3.26$&$2.66$     &$1.90$&$1.78$\\
\hline
\end{tabular}
\end{center}
\label{tab:vetocuts}
\end{table} 

\begin{figure}
  \resizebox{\hsize}{!}{\rotatebox{-90}{\includegraphics{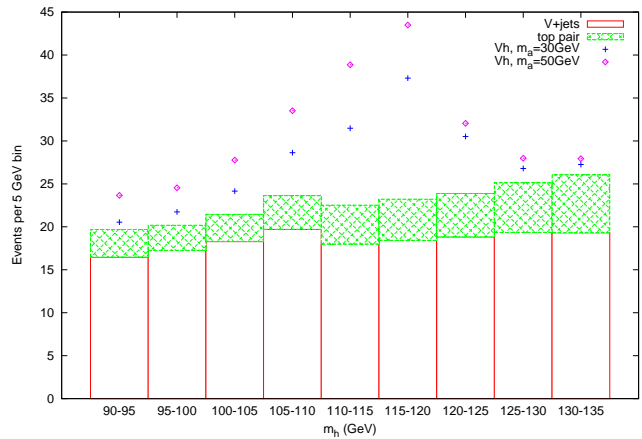}}}
  \caption{Expected events (per $5$ GeV bin) versus the reconstructed Higgs mass given $100$ fb$^{-1}$ of data. Two different signal cases are presented, with $m_a = 30$ GeV and $m_a = 50$ GeV. Both sets of points represent a model where $a \rightarrow b\bar{b}$ is the only decay channel. The data represents events which make it through all cuts discussed in this section except the cut on $m_h$ (jet veto, $\geq3$ subjets, and $\geq2$ $b$-tagged subjets). }  
\label{fig:hmassbtags}
\end{figure}

With these kinematic cuts in place, we turn our attention to $b$-tagging of the constituent subjets. Our model of tagging efficiencies and purities is relatively simple: displaced vertices arising from $b$ and $c$ hadrons (culled from Monte Carlo level information) are assigned to the nearest (in the $\Delta R$ measure) subjet, as long as that subjet is less than a fixed $\Delta R_{max}$ (set here to $0.2$) from the vertex. Subjets with assigned vertices are tagged with fixed efficiencies ($\epsilon_{b}$,$\epsilon_{c}$) given $b$ and $c$ vertices, respectively. If a single subjet is assigned multiple vertices then each vertex provides an independent tagging probability (e.g. the tagging probability of a subjet assigned two $b$ vertices is $1-(1-\epsilon_{b})^2$). Similarly, the tagging probability on different subjets is assumed to be independent. If a subjet is not assigned any vertices then it may still be tagged with efficiency $\epsilon_l$. We use the values $\epsilon_{b}=0.4$, $\epsilon_{c}=0.1$, $\epsilon_{l}=0.02$. We require that at least two subjets are $b$-tagged. With these efficiencies and requirements, the $V+jets$ background cross-section is reduced from $138.6$ fb to $0.93$ fb and the $t\bar{t}$ background from $7.86$ fb to $0.21$ fb. The impact of $b$-tagging on our signal is summarized in Table \ref{tab:heavyresults}. We also present, in Figure \ref{fig:hmassbtags}, the reconstructed uncleaned Higgs mass (as in Figure \ref{fig:cleanvunclean}) for signal and backgrounds after three cuts are taken into account: the jet veto, the splitting of the candidate Higgs into $3+$ subjets and the $b$-tagging $2+$ subjets.  

\begin{table}
\caption{Expected number of signal events (given $100$ fb$^{-1}$ of data) after $b$-tagging and all cuts, along with expected signal to background ratio and significance (as before). } 
\begin{center}
\begin{tabular}{|c|c|c|c|}
\hline
$m_a($GeV$)$&$Signal$&$s/b$&$\sigma$\\
\hline
$20$&$1.9$&$2\%$&$0.18$\\
\hline
$30$&$37.4$&$33\%$&$3.49$\\
\hline
$40$&$63.1$&$55\%$&$5.89$\\
\hline
$50$&$61.0$&$53\%$&$5.69$\\
\hline
\end{tabular}
\end{center}
\label{tab:heavyresults}
\end{table} 

\section{Conclusion}
By restricting our search to a very specific part of phase space (high $p_T$ associated Higgs production), it is possible to strongly reduce the SM background while only weakly reducing the Higgs signal. The high $p_T$ threshold on the vector boson leads to a signal sample of high $p_T$ Higgs bosons, which means that the decay products of the Higgs, under any physics model, tend to be collimated in a relatively small region in the detector. Using analysis of the subjet structure present due to cascade Higgs decays through a light scalar we've demonstrated the feasibility of a flavor-independent search for scalar masses from $\sim 15 - 30$ GeV with $\sim 100$ fb$^{-1}$.  This, for example, is the broadest parameter space suggested by any search strategy thus far presented for the decay $h \rightarrow 2a\rightarrow 4g$ and thus may be a viable search at the LHC. If cascade Higgs decays proceed primarily to $b$ quarks then flavor tagging can increase the significance of the search dramatically. For the high range of scalar masses, $35$ GeV $\leq m_a \leq \frac{m_h}{2}$ the options are more limited; assuming reasonable $b$ tagging efficiency and purity, discovery significance can also be achieved with $\sim 100$ fb$^{-1}$.  


\begin{thebibliography}{1}

\bibitem{ewwg}
See http://lepewwg.web.cern.ch/LEPEWWG/.

\bibitem{Chang:2008cw}
  S.~Chang, R.~Dermisek, J.~F.~Gunion {\it et al.},
  Ann.\ Rev.\ Nucl.\ Part.\ Sci.\  {\bf 58}, 75-98 (2008).
  [arXiv:0801.4554 [hep-ph]].

\bibitem{LHCInvH}
 O.J.P.Eboli, D.Zeppenfeld
 Phys.\ Lett.\  B {\bf 495} (2000)
 arXiv:hep-ph/0009158v1.

\bibitem{Schael:2006cr}
  S.~Schael {\it et al.} [ ALEPH and DELPHI and L3 and OPAL and LEP Working Group for Higgs Boson Searches Collaborations ],
  Eur.\ Phys.\ J.\  {\bf C47}, 547-587 (2006).
  [hep-ex/0602042].

\bibitem{Lisanti:2009vy}
  For a nice discussion, see M.~Lisanti, J.~G.~Wacker,
  Phys.\ Rev.\  {\bf D81}, 096005 (2010).
  [arXiv:0911.1997 [hep-ph]].

\bibitem{Chang:2005ht}
  S.~Chang, P.~J.~Fox, N.~Weiner,
  JHEP {\bf 0608}, 068 (2006).
  [hep-ph/0511250].

\bibitem{Bellazzini:2009kw}
  B.~Bellazzini, C.~Csaki, A.~Falkowski {\it et al.},
  Phys.\ Rev.\  {\bf D81}, 075017 (2010).
  [arXiv:0910.3210 [hep-ph]].

\bibitem{Ellwanger:2003jt}
  U.~Ellwanger, J.~F.~Gunion, C.~Hugonie {\it et al.},
  [hep-ph/0305109];
  U.~Ellwanger, J.~F.~Gunion, C.~Hugonie,
  JHEP {\bf 0507}, 041 (2005).
  [hep-ph/0503203].

\bibitem{Carena:2007jk}
  M.~Carena, T.~Han, G.~-Y.~Huang {\it et al.},
  JHEP {\bf 0804}, 092 (2008).
  [arXiv:0712.2466 [hep-ph]].
    
\bibitem{Butterworth:2008iy}
  J.~M.~Butterworth, A.~R.~Davison, M.~Rubin {\it et al.},
  Phys.\ Rev.\ Lett.\  {\bf 100}, 242001 (2008).
  [arXiv:0802.2470 [hep-ph]].

\bibitem{Butterworth:2002tt}
  J.~M.~Butterworth, B.~E.~Cox, J.~R.~Forshaw,
  Phys.\ Rev.\  {\bf D65}, 096014 (2002),
  [hep-ph/0201098];
  J.~M.~Butterworth, J.~R.~Ellis, A.~R.~Raklev,
  JHEP {\bf 0705}, 033 (2007),
  [hep-ph/0702150 [HEP-PH]].

\bibitem{Falkowski:2010hi}
  A.~Falkowski, D.~Krohn, L.~-T.~Wang {\it et al.},
  [arXiv:1006.1650 [hep-ph]].

\bibitem{Chen:2010wk}
  C.~-R.~Chen, M.~M.~Nojiri, W.~Sreethawong,
  JHEP {\bf 1011}, 012 (2010),
  [arXiv:1006.1151 [hep-ph]].

\bibitem{Bellazzini:2010uk}
  B.~Bellazzini, C.~Csaki, J.~Hubisz {\it et al.},
  [arXiv:1012.1316 [hep-ph]].

\bibitem{mcevoy-thesis}
~From M.~McEvoy, ``Non-standard Higgs Decays at the LHC'', PhD. Thesis, Mar 2010, D.~E.~Kaplan, advisor.

\bibitem{MadGraph}
  T. Stelzer and W.F. Long
  Phys. Commun. {\bf 81} (1994)
  arXiv:hep-ph/9401258.

\bibitem{MadEvent}
  F. Maltoni and T. Stelzer
  JHEP {\bf 0302} (2003)
  arXiv:hep-ph/0208156.

\bibitem{Sjostrand:2006za}
  T.~Sjostrand, S.~Mrenna and P.~Skands,
  JHEP {\bf 0605}, 026 (2006)
  [arXiv:hep-ph/0603175].

\bibitem{FastJet}
  Matteo Cacciari, Gavin Salam and Gregory Soyez,
  Based on work by Matteo Cacciari and Gavin Salam
  Phys.Lett.B {\bf 641} (2006)
  arXiv:hep-ph/0512210v2

\bibitem{AngularResolution}
 The ATLAS Collaboration,
 CERN-OPEN-2008-020 pp 368-392

\bibitem{CMScal}
  The CMS collaboration
  CERN/LHCC 2006-001 (2006)
  146-224
  
\bibitem{Dokshitzer:1997in}
  Y.~L.~Dokshitzer, G.~D.~Leder, S.~Moretti {\it et al.},
  JHEP {\bf 9708}, 001 (1997),
  [hep-ph/9707323];
  M.~Wobisch, T.~Wengler,
  [hep-ph/9907280].

\bibitem{BoostedTops}
 David E. Kaplan, Keith Rehermann, Matthew D. Schwartz, and Brock Tweedie,
 Phys.Rev.Let. {\bf 101} (2008)
 arXiv:0806.0848v2 [hep-ph].

\bibitem{Ellis:2009su}
  S.~D.~Ellis, C.~K.~Vermilion, J.~R.~Walsh,
  Phys.\ Rev.\  {\bf D80}, 051501 (2009).
  [arXiv:0903.5081 [hep-ph]].

\bibitem{:2001yb}  See 
  [ LEP Higgs Working Group for Higgs boson searches Collaboration ],
  [hep-ex/0107034], for the ``flavorless'' Higgs search.  Supersymmetry searches such as
  S.~Schael {\it et al.} [ ALEPH Collaboration ],
  Eur.\ Phys.\ J.\  {\bf C49}, 439-455 (2007),
  [hep-ex/0605079] and
  G.~Abbiendi {\it et al.} [ OPAL Collaboration ],
  Phys.\ Lett.\  {\bf B545}, 272-284 (2002),
  [hep-ex/0209026], may also be relevant for the $Zh$ production when the $Z$ decays invisibly.





































%


























  


\end{thebibliography}
\end{document}